# Electromotive force and magnetization process of a superconducting traveling-wave flux pump

Wei Wang[1,*], Jiafu Wei[1], Chenghuai Wu[1], Guangtong Ma[2], Hong Li[1], Hanxin Ye[1], Yuntian Zhang[1]

[1] *College of Electrical Engineering, Sichuan University, Chengdu, Sichuan, 610065, China.*
[2] *Applied Superconductivity Laboratory, State Key Laboratory of Traction Power, Southwest Jiaotong University, Chengdu, Sichuan 610031, China.*

Understanding and controlling the motion of superconducting vortices has been a key issue in condensed matter physics and applied superconductivity. Here we present a method for macroscopically manipulating the vortices based on travelling wave flux pump to accurately output industrial-scale DC current into high-temperature superconducting (HTS) magnets. DC magnetic fields are used to adjust the polarity of the vortices and thus modulate the direction of the output current, which demonstrates that the DC current of the flux pump originates from the motional electromotive force ( *e.m.f.* ) other than the induced *e.m.f.*. In addition, applying different strengths of DC fields can modulate the magnitude of the output current. Further numerical simulation suggests how the flux inside the superconducting tape is controlled by different applied fields. We build a controlled flux flow model to correctly explain the behavior of vortices controlled by the flux pump, and how the motional *e.m.f.* is created by manipulating the vortices. Based on the method, we achieve high precision regulation of output current using adaptive control of the DC magnetic field, allowing the flux pump to output DC current just as accurate as a typical commercial power supply. This work advances the technic for macroscopic manipulation of vortices.

* Corresponding author. Email: weiwangca283@gmail.com (Wei Wang)

## I. INTRODUCTION

When a type-II high-temperature superconductor is in the mixed state, magnetic field can penetrate into the interior of the superconductor in the form of vortices. The vortex motion can be experimentally controlled by transporting current [1], magnetic field [2], and temperature [3],and so on. In recent years, it has been possible to handle one or several vortices precisely with external magnetic field by using tiny magnets [4] and scanning probe techniques [5-7], temperature gradient generated by laser beam [8], and so on, which helped to advance the application of fluxtronics [9,10]. On the other hand, Manipulating millions of clustered vortices could lead to an powerful application that could shake up the superconducting industry [11]— a contactless DC power source for charging HTS magnets based on HTS flux pumps [12,13]. A flux pump can directionally manipulate the movement of vortices by applying a spatially travelling wave magnetic field [14], and output DC current into an HTS magnet at a kilo-ampere level [15,16], an innovative power transmission similar to coupled magnetic resonance [17]. Comparing to the conventional commercial DC power sources, flux pumps reduce the size, weight, energy consumption and cost by at least one order, meanwhile, it discards the harmful current lead and its inherited heat load, which make it become indispensable for the next generation HTS magnets in extensive applications.

The initial prototype to induce DC current by controlling the directional movement of vortices was performed by Giaever in the 1960s (which was called "DC transformer") [2,18,19]. In that system, two layers of superconducting film were separated by a silicon oxide insulation layer, and vortices in two films are coupled and move in synergy when transporting current into one film, and then generate a DC current in the other film due to the directional movement of the vortices. This effect is different from the "ratchet effect" [20,21] because the film in DC transformer does not require the asymmetric potential, but instead relies on the external driving force to guide directional movement of vortices. Similar coupling systems can be found in two-dimensional electron gases based on Coulomb drag [22], mesoscopic dynamo based on dynamic heat drag [23,24], and importantly, the HTS flux pump based on magnetic coupling [11,25]. In an HTS flux pump system, the external travelling magnetic field is applied to replace the superconducting film with transporting current in the DC transformer, and the vortex motion is controlled by magnetic coupling force [11,13]. However, the output current of a flux pump is dramatically larger by three orders compared with that of a DC transformer, which can meet the charging demand of HTS magnets.

The motion of vortices may generate flux flow resistance and energy dissipation, due to the movement of normal state electrons inside the magnetic core and outside electrons [26], and the damped equation of a vortex can be written as [19,26]

$$\eta v = F_d - F_p \quad (1)$$

where $\eta$ is Bardeen–Stephen friction viscosity [1], $v$ is the velocity of the vortex, $F_d$ is the dragging force, $F_p$ is the pinning force, and the repulsive force between vortices and other resistances are ignored [27] in order to simplify the motion of vortices in Eqn.(1). When the dragging force $F_d$ is larger than the resistance force $F_p$, the dragging force may drive the vortex to move. During the dragging process, if a large number of vortices are manipulated to move in one direction, it can generate a large DC output according to $\vec{E} = N\vec{\phi}_0 \times \vec{v}$, where $N$ is the effective number of vortices and $\phi_0$ is the flux quantum [19]. Actually the existing numerical methods such as $H$-formulation [28] and $T$-$A$ formulation [29], which are widely used in superconducting electromagnetism, cannot directly simulate the DC output by travelling wave flux pumps [30,31]. This is because most simulation objects in superconducting electromagnetism usually do not produce the physical effect caused by the motion of mesoscopic vortices, and naturally ignore the key motional electromotive force ( $e.m.f.$ ), that is, the second term of the $e.m.f.$ equation [13,32],

$$e.m.f. = -\partial_t \vec{A} + \vec{v} \times \vec{B} \quad (2)$$

where the first term is the induced $e.m.f.$ from Faraday's law of induction, $\vec{A}$ is the magnetic vector potential, i.e. $\vec{B}=\nabla\times\vec{A}$. The second term is the motional $e.m.f.$, derived from the Lorentz transformation due to the relative motion between the charge and the magnetic field (named as the "Lorentz term"). As a result, the induction term ( $-\partial_t\vec{A}$ ) alone cannot thoroughly reflect the nature of macroscopic controlled motion of vortices — an unreported special case of "exceptions to the flux rule" [32], as a result of motional $e.m.f.$ during the coupled flux flow of vortices [11,13].

In this work, we show the tremendous application value of macro-manipulating vortices performed by a linear-motor flux pump [33,34]. We demonstrate the experimental evidences that the DC output voltage of the flux pump is originated from motional *e.m.f.* (the Lorentz term) of directionally moving vortices, other than the induced *e.m.f.* (induction term). By modulating the polarization of superconducting vortices with an external DC bias magnetic field, the DC output of the flux pump can be accurately controlled. We provide a numerical simulation to reveal the internal coupled flux flow inside the YBCO stator, and build a magnetization model to elucidate the origin of the motional *e.m.f.* during the coupled flux flow process. In the end, we demonstrate the tremendous application value of using the external magnetic field to manipulate millions of vortices — by using adaptive control [35], the output DC current is accurately controlled, to be compared to a conventional commercial power supply based on power electronics, which enables the flux pump to be applied to some highly demanding fields such as charging MRI magnets.

## II. RESULTS

### A. The physical origin of "electromotive force" of flux pump

In the last decade, the major reason for the confusion of travelling wave HTS flux pump is that, it's not clear how a DC "electromotive force" is generated by a travelling magnetic field. The *e.m.f.* is only derived from two components as Eqn.(2), where the first is the induced *e.m.f.* $(-\partial_t \vec{A})$, the second is the motional *e.m.f* $(\vec{v} \times \vec{B})$. We demonstrate experimentally to show that the *e.m.f.* of the travelling wave flux pump is the direct result of the motional *e.m.f.* $(\vec{v} \times \vec{B})$ rather than the induced *e.m.f.* $(-\partial_t \vec{A})$.

Travelling wave flux pumps mainly include HTS dynamo [12] and linear-motor flux pump [34], both of which generate adjustable spatial magnetic field, expressed by one dimensional wave function [13]:

$$B_y(x,t) = B_a \sin(kx + \omega t) + B_{bias} \quad (3)$$

where $B_a \sin(kx+\omega t)$ is the AC travelling wave, $B_{bias}$ is an imposed DC bias field.

The demonstrative experiments are carried out on a linear-motor flux pump, as shown in Fig.1(a). This device produces the AC and DC fields by separate coils, to be controlled independently. The imposed DC bias field $B_{bias}$ is a static field, which doesn't take part in the magnetic induction, such as it is "invisible" in the induction law after time-derivative. In this work, we demonstrate the following three experimental cases, i.e. keeping the AC field component $B_a \sin(kx+\omega t)$ the same, but change the DC bias fields $B_{bias}$ in three different values, such as $B_{bias} = 0$, $B_{bias} = +B_d > 0$, and $B_{bias} = -B_d < 0$. The travelling magnetic waves for the three cases are demonstrated in Fig.1(b), which are expressed as:

Case (I):      $B_{y1}(x,t) = B_a \sin(k_0 x + \omega_0 t)$     (4)

Case (II):     $B_{y2}(x,t) = B_a \sin(k_0 x + \omega_0 t) + B_d$     (5)

Case (III):    $B_{y3}(x,t) = B_a \sin(k_0 x + \omega_0 t) - B_d$     (6)

We can freely adjust the bias field $B_{bias}$ to an arbitrary value. For example, we choose $B_d = B_a$ to just fully bias the applied waves above or below zero axis. Since the time-derivatives of Eqn.(4)-(6) are entirely the same, the induced *e.m.f.* in Eqn.(2) are entirely the same, i.e. $-\partial_t \vec{A}_1 = -\partial_t \vec{A}_2 = -\partial_t \vec{A}_3$. If the *e.m.f.* is a result of magnetic induction, then the flux pump's DC outputs should be entirely the same in the three cases.

Fig.1(c) shows the measured "pumped" current inside the YBCO coil. Surprisingly, the three experiments results in three different pumped currents, such as: in Case (I) when $B_{bias} = 0$, there is zero pumped current, i.e. $I_{pump} = 0$ A; in Case (II) when $B_{bias} > 0$, there is negative pumped current, i.e. $I_{pump} < 0$ A; in Case (III) when $B_{bias} < 0$, there is positive pumped current, i.e. $I_{pump} > 0$ A. For each given value of $B_d$, the pumped currents are symmetric for Case (II) and (III), e.g., in the case $B_d = B_a$, Case (II) obtains $I_{pump} = -69$ A, while Case (III) obtains $I_{pump} = 69$ A. In other words, this static field $B_{bias}$ had surprisingly caused a variation of output current up to 138 A. Those experiments strongly suggest that, the *e.m.f.* of the travelling wave flux pump is not a result of induction, as the induction terms $-\partial_t \vec{A}$ are entirely the same for the three Cases. Based on Eqn.(2), the only possible *e.m.f.* for the flux pump is from the motional *e.m.f.*, i.e. the Lorentz term $\vec{v} \times \vec{B}$.

In addition to the above bidirectional regulation, a tremendous important experimental phenomenon is that, the output DC current of the flux pump can be continuously adjusted by the DC bias field, while the AC magnetic field was kept the same. The output of the flux pump responds immediately to the change of $B_{bias}$. As can be seen in Fig.1(c), when $B_d$ increases from 0 to $B_a$, the pumped current increases almost linearly, which realizes the free adjustment of the flux pump's output. As a result, by controlling the DC bias magnetic field, we realize the continuous regulation of the DC output of a flux pump as well as the reverse regulation, which enables the excellent controllability of the flux pump as a superconducting DC power source.

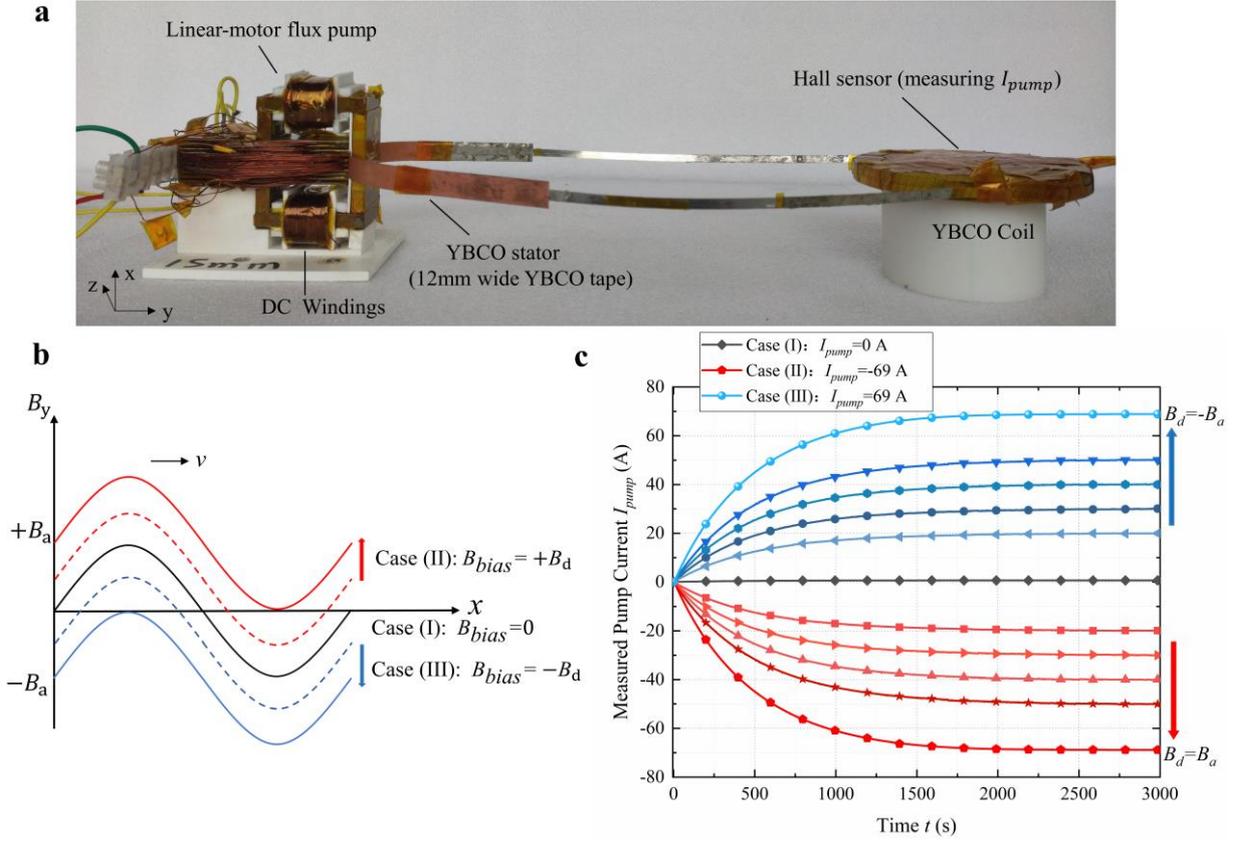

**Fig. 1 The three cases demonstrated the impact of DC bias field on flux pump's DC output current. a** the experimental setup based on a linear-motor flux pump and a closed-loop YBCO coil, while the pumped currents $I_{pump}$ were measured by a Hall sensor placed at the center of the coil. **b** the applied magnetic waves in the three cases, which have the same AC component and magnetic induction, but different DC bias fields which can be freely adjusted. **c** the measured pumped current $I_{pump}$ in the three different cases, which resulted in three different pumped currents, i.e. in Case (I), when $B_{bias}$=0, $I_{pump}$=0 A; in Case (II), when $B_{bias} > 0$, $I_{pump} < 0$; in Case (III), when $B_{bias} < 0$, $I_{pump} > 0$.

**B. Controlled flux flow model**

The above experiments strongly indicate that the travelling wave flux pump is an unrevealed case of "exceptions to the flux rule" as in *Feynman's Lectures on Physics* [32], but entirely different from existing cases, such as there is no moving magnet in above experiments. Typical example of "exceptions to the flux rule", such as homopolar generator, which generates motional *e.m.f.* from a rotating magnet [36]. To explain how the motional *e.m.f.* was generated by the travelling wave flux pump, here we build a controlled flux flow model, which elucidates why a static field have the power to regulate the flux pump's *e.m.f.*.

We start with a numerical model based on the *H*-formulation [28] and *E-J* power law with constant critical current density $J_C$, as described in the Methods. We have previously shown that, this numeral method can accurately simulate the flux flow process inside the YBCO film under travelling waves [11,14,37]. Now we use this model to simulate the flux flow inside the 12 mm wide YBCO stator in the experiments shown in Fig.1(a). The flux flow processes and current distributions in the YBCO stator for the three experiments are shown in Fig.2, i.e. $B_{bias} = 0$, $B_{bias}=+B_a$, $B_{bias} = -B_a$ respectively. Since $B_{bias}$ doesn't take part in the magnetic induction and constant $J_C$ is considered, it is not surprising that the induced current distributions are entirely the same for the three cases. Since the induced *e.m.f.* is derived from the superconducting *E-J* relationship, thereby, the induced *e.m.f.* are also the same for the three cases. That is to say, the *e.m.f.* of the travelling wave flux pump is clearly not a result of magnetic induction, and the *E-J* power law with constant $J_C$ might fall short to predict travelling wave flux pump's DC output.

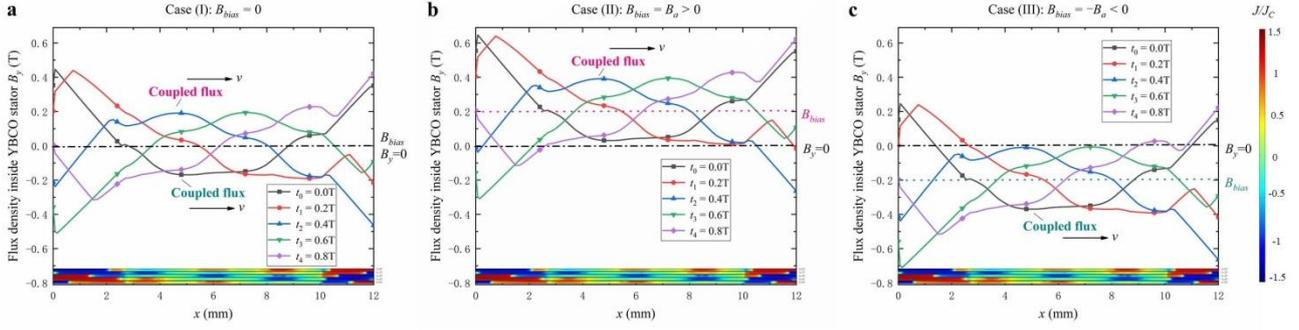

**Fig. 2** The simulation results of internal flux flow and current distributions inside the YBCO stator in the three cases based on the *H*-formulation and *E-J* power law with constant $J_C$. **a** the penetrated flux inside the YBCO stator in Case (I), below are the current distributions at each time step; **b** Case (II); **c** Case (III).

As the travelling magnetic waves sweep across the 12 mm YBCO stator, as shown in Fig.2, we unveil a clear "magnetic coupling" phenomenon [11] in the central region, i.e. some flux is coupled and moving in accordance with the magnetic poles. To understand the magnetic coupling process, we build a controlled flux flow model based on the consumption of constant $J_C$. The flux flow process is as following:

(i) as shown in Fig.3(a)-(b), when a magnetic pole moves across the left edge of the YBCO stator, based on the Bean model [38], the oscillating field will induce a magnetic moment (circulating current) on the left edge, with the induced current density equals to $J_C$. In classic magnetism, when a magnetic moment is exposed to an inhomogeneous field such as the magnetic pole, it will endure unbalanced Lorentz force as shown in Fig.3(b). This Lorentz force will "drag" the induced magnetic moment to move in synergy with the applied pole.

(ii) As shown in Fig.3(c)-(d), the magnetic pole "couples" part of the flux to move across the central region of the 12 mm wide YBCO stator, Fig.2 suggests the "coupled" magnetic moment has a smaller current density $J_m$ compared with the critical current $J_C$. The "coupling force" between the coupled current loop and the applied magnetic pole was calculated as:

$$F_{couple} = J_m \left[ \int_{x_2}^{x_3} B_y(x)dx - \int_{x_1}^{x_2} B_y(x)dx \right] \quad (7)$$

where $F_{couple}$ provides the dragging force, such as $F_d$ in Eqn.(1). The coupled current loop is comprised of a cluster of superconducting vortices, in order to force the coupled vortices to move in synergy, $F_{couple}$ must be larger than the pinning force $F_p$ of the vortices cluster, i.e. $F_{couple} > F_p$ in Eqn.(1). The moving speed $\vec{v}$ of this controlled flux flow, however, equals to the travelling speed of the magnetic pole, i.e. $v = \lambda f$, where $\lambda$ is the spatial wavelength, $f$ is the applied frequency.

During this directional flux flow, the moving vortices give rise to a motional *e.m.f.* [26,39] based on Eqn.(2). Since $\vec{v}$ here is the relative speed of the moving vortices to free charges, thereby, during this coupled flux flow, the output DC voltage $\vec{V}$ is expressed as:

$$\vec{V} = l_{eff} \bar{\vec{B}} \times \vec{v} \quad (8)$$

where $\bar{B}$ is the averaged flux density of coupled vortices, $l_{eff}$ is the effective length along the YBCO stator.

Thereby, despite there is no visible moving magnets exist in the experiments shown in Fig.1(a), the coupled moving of the reference frame of superconducting vortices can still generate a motional *e.m.f.*, those vortices can be considered as tiny magnets exist in mesoscopic scale. In the whole process while the magnetic pole sweeps across the YBCO stator, only the central region with directional flux flow give rise to DC output voltage, based on Eqn.(8).

(iii) As shown in Fig.3(e)-(f), the magnetic pole moves across the right edge, which induces screen current on the right edge of the YBCO stator.

In the whole magnetization process, the YBCO stator can be divided into three regions, i.e. two oscillating regions on the edges to induce screen currents and magnetic moments, the coupled flux flow region in the central region of the YBCO stator, while only the coupled flux flow region outputs DC voltage based on Eqn.(8). Thereby, the YBCO stator is recommended to be sufficiently wide, to increase the coupled flux flow region in the middle.

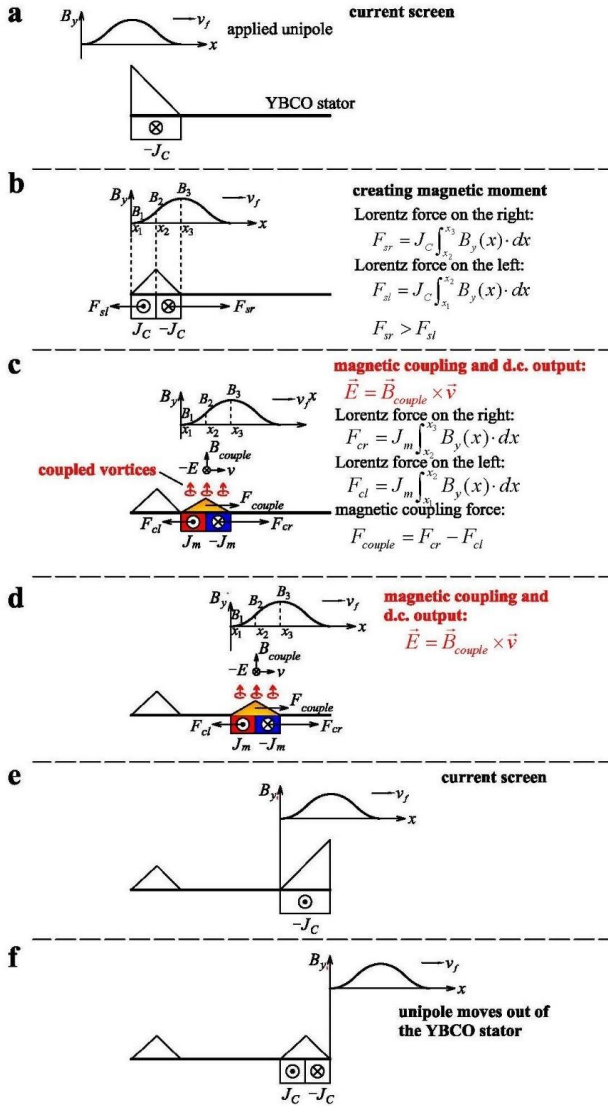

**Fig. 3 The proposed flux flow model when a magnetic pole moves across the YBCO stator.** **a** the peak of the magnetic pole moves at the left edge of the YBCO stator, a screen current is induced; **b** the rear of the magnetic pole move across the left edge, induces a magnetic moment, the magnetic pole provides unbalanced Lorentz force to "drag" the magnetic moment towards the moving direction; **c** and **d** the unipole "couples" part of the magnetic moment to move across the YBCO stator, which outputs motional *e.m.f.* based on Eqn.(7); **e** and **f** the magnetic pole induces screen current on the right edge of the YBCO stator.

## C. Modulation of vortices by a static DC magnetic field

Based on the controlled flux flow model, we now explain why the DC magnetic field $B_{bias}$ has the power to regulate flux pump's DC output as experimentally demonstrated in Fig.1. We now consider the AC travelling wave contains a pair of positive and negative magnetic poles in one period. By using the mean method in statistics, we assume that, during the coupled flux flow, the positive pole couples a number of $N_1$ positive vortices, each with a flux quantum $\phi_0$; while the negative pole couples a number of $N_2$ negative vortices, each with a flux quantum $-\phi_0$. Based on Eqn.(8), it give rise to a total motional *e.m.f.* $\vec{E}$ in one period, expressed as:

$$\vec{E} = \frac{(N_1\phi_0\vec{e}_y - N_2\phi_0\vec{e}_y)}{S_{eff}} \times \vec{v} = -\frac{(N_1 - N_2)\phi_0 v_0}{S_{eff}}\vec{e}_z \qquad (9)$$

while considering $\vec{v} = v_0\vec{e}_x$ as shown in Fig.1(b), $S_{eff}$ is the effective area of coupled vortices.

In Case (I), as shown in Fig.4(a), since the AC travelling wave is symmetric, the positive and negative poles couple positive and negative vortices that are equal in number, i.e. $N_1 = N_2$, which results in zero DC output based on Eqn.(9), as experimentally demonstrated in Fig.1(c).

In the Case (II), as shown in Fig.1(b), the AC travelling wave is "lifted" by the DC bias field, i.e. $B_{bias} = +B_d > 0$, the positive and negative poles are asymmetric, results in unbalanced numbers of coupled positive and negative vortices, i.e. $N_1 > N_2$, resulting in a negative DC output, i.e. $\vec{E} < 0$ based on Eqn.(9). It can be seen that, by gradually increasing $B_{bias}$, the value of $(N_1 - N_2)$ in Eqn.(9) is gradually increased, while the maximum value $(N_1 - N_2)_{max}$ occurs at $B_{bias} = B_a$, as shown in Fig.4(b), where there is just no negative pole, resulting in $N_2 = 0$ and $N_1$ reaches maximum, which gives rise to a maximum DC output based on Eqn.(9). This theoretical prediction is supported by another reported experiment [40]. In other words, by modulating the value of $B_{bias}$ between 0 and $B_a$, the DC output of the flux pump can be regulated to any value between 0 and $E = -|E_{max}| = -|N_1 - N_2|_{max}\phi_0 v_0$, as was demonstrated in Fig.1(c).

Case (III) is explained similarly as in Case (II), the DC bias field $B_{bias} = -B_d < 0$ has "lowered" the AC travelling wave, resulting in $N_1 < N_2$ and a positive DC output $\vec{E} > 0$, based on Eqn.(9). That is to say, by reversing the direction of the DC bias field, the flux pump's DC output can be effectively reversed. This phenomenon is also experimentally verified for the HTS dynamo [35]. Similarly, as shown in Fig.4(c), the maximum DC output voltage occurs at $B_{bias} = -B_a$, while there is just no positive pole. By modulating the value of $B_{bias}$ between 0 and $-B_a$, the DC output of the flux pump can be regulated to any value between 0 and $E = +|E_{max}| = +|N_1 - N_2|_{max}\phi_0 v_0$, as was demonstrated in Fig.1(c).

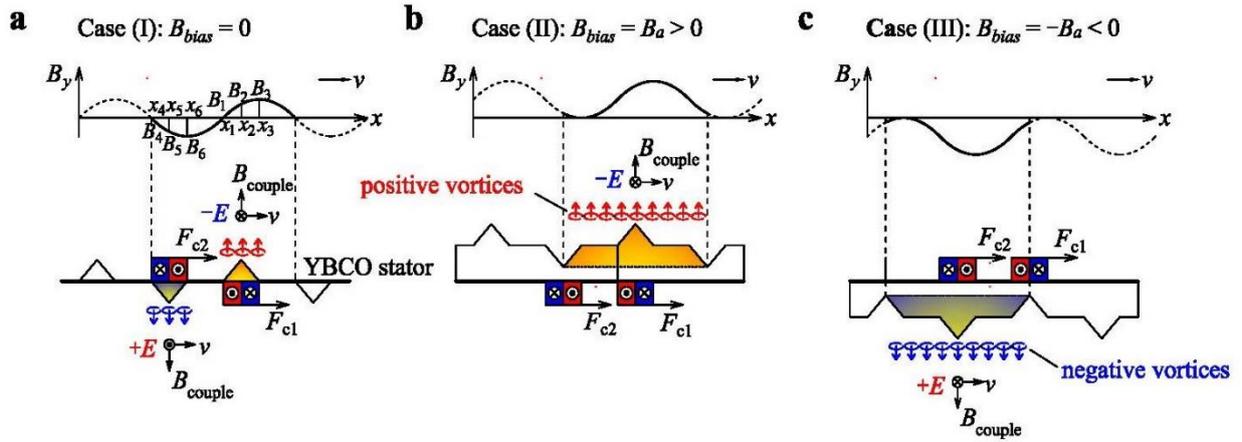

**Fig. 4 The role of DC bias field on flux pump's DC output.** To help understand, the YBCO stator is considered to be very wide, while the penetrated flux is periodical. **a** in Case (I) $B_{bias}=0$ the AC travelling wave has both positive and negative poles, which couple positive and negative vortices of the same number, thus output zero DC electric field; **b** in Case (II) $B_{bias}>0$, the DC bias field modulates all the coupled vortices to positive direction, which output negative DC electric field $-E$; **c** in Case (III) $B_{bias}<0$, the DC bias field modulates all the coupled vortices to negative direction, which output positive DC electric field $E$.

The directional relationship between $\vec{V}$, $\vec{B}$ and $\vec{v}$ as in Eqn.(8) is experimentally verified with an array of Hall sensors.

The above trait enables the flux pump as a bidirectional DC power source for the HTS magnets, i.e. by modulating the DC bias field from $-B_a$ to $+B_a$, the DC output can be continuously adjusted from $+|E_{max}|$ to $-|E_{max}|$, while the pumped current can also be continuously adjusted from $+|J_{max}|$ to $-|J_{max}|$ as demonstrated in Fig.1(c). The change of flux pump's DC output is spontaneous as the DC field changes, which enables a powerful way for accurate current control as will demonstrated in the following.

**D. Enable high accurate current control of a flux pump**

Based on the above discussion, by switching the DC bias field $B_{bias}$ between 0 and $B_d \neq 0$, the DC output $\vec{E}$ can be switched between 0 and $-(N_1 - N_2)\phi_0 v_0 \vec{e}_z$. Since the DC coil can be switched on and off at a very fast frequency, with the help of a feedback control program (shown in Methods), the output current of the flux pump can be controlled very accurately.

One example is demonstrated in Fig.5. In each case, we preset a target load current $I_L$ and allowed current fluctuation $\Delta I$, e.g., we preset $I_L = 30\ A$ and $\Delta I = 0.15\ A$. As shown in Fig.5, after the load current had been pumped to 30.075 A, the DC coil was switched off, while the load current decays slowly due to the soldering resistance. After it decays to 29.925 A, the DC coil was switched on again to increase the load current. With this feedback control, as shown in Fig.5, the load current can be controlled very accurately to any preset load current $I_L$ and accuracy $\Delta I$. This essentially enable the flux pump as a very accurate bidirectional power source for the superconducting load, as comparable as a high accuracy conventional power supply based on power electronics, however, without current leads and has much lower energy consumption. This method enables the flux pump as a promising current source candidate for the HTS MRI magnets, which requires very stable current at the fluctuation of part per million (ppm).

**III. DISCUSSION**

The above experiments demonstrate the significant application value of macroscopic regulation of vortices, which results in large DC current output and high accurate current control. This simple and efficient method of manipulating vortices based on external magnetic field can greatly expand the applied research. In fact, very few works have been reported in the field of superconductivity to generate electricity based on manipulating vortices, that can have practical value in applied physics, except for flux pumps. In this work, we have demonstrated the use of a static DC magnetic field to continuously change the flux pump's DC output from $-69\ A$ to $+69\ A$, without change the induction term $-\partial_t \vec{A}$, which had verified that the DC output of flux pump is originated from the motional *e.m.f.* due to the control flux flow of superconducting vortices, thus has verified the flux pump as an unrevealed special case of "exception to the flux rules", while the role of the DC bias field is essentially to change polarities of the coupled vortices. Before this work, it has not been realized that the polarity of mesoscopic superconducting vortices may have a visible impact on the macroscopic world. Especially, in the present macroscopic electromagnetic modeling of type-II superconductors, a static magnetic field is deleted in the induction law with superconducting *E-J* relationship considering constant $J_C$. This work suggested that, the motional *e.m.f.* from the vortices flow, i.e. Eqn.(8), may play a vital role in certain applications of superconductivity, however, might have been missed out in our present superconducting constitutive relationship (see Methods), which suggests an in-depth modification is might at the doorstep.

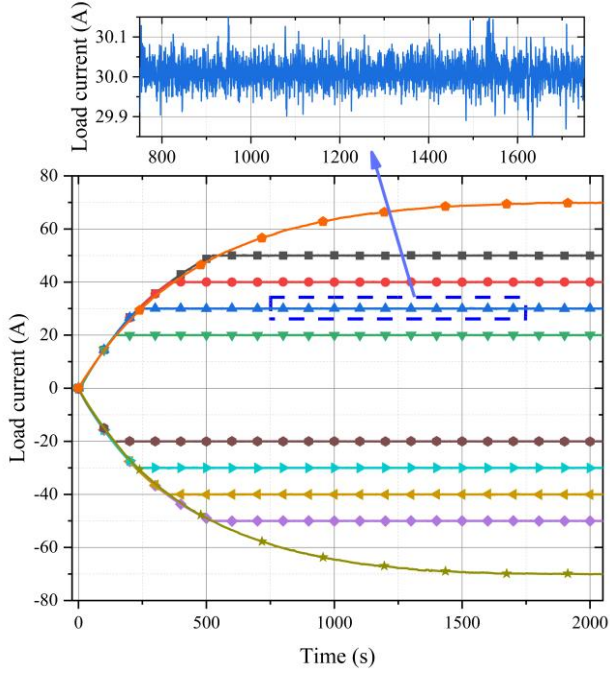

**Fig. 5 Arbitrary adjustment of the flux pump's output current.** By adjusting the on/off state of the DC magnetic field, we can arbitrarily adjust the output current of the flux pump, so that the flux pump can output any preset value below the maximum current. The accuracy of the output current is also guaranteed with a fluctuation of 0.15 A based on feedback control.

Subsequently, we build a controlled flux flow model to illustrate how an applied magnetic pole induces a magnetic moment on the edge of a 12 mm wide YBCO stator, and drags it through the stator. We also build the quantitative Eqn.(7) to evaluate the coupling force $F_{couple}$, which provides the dragging force $F_d$ in the basic vortex dynamics Eqn.(1). Very few work has reported to incorporate the magnetic coupling force into the vortex dynamics, the main reason is that, for most cases, the applied magnetic field is reasonably homogenous at the region the superconductor, while the coupling force due to magnetic gradient can be reasonably ignored. For instance, the Bean model [38] considers a homogenous applied field, based on Eqn.(7), it provides a balanced Lorentz to the induced magnetic moment on the edge of the superconductor, while the magnetic coupling shall not exist. However, this is clearly not the case for the travelling wave flux pump, as the applied field in quite inhomogeneous for the superconducting stator, which leads to non-zero coupling force for the existing vortices, i.e. $F_{couple} > 0$. Consequently, we have found that, by increasing the field inhomogeneity, such as by increasing the field amplitude $B_a$ and shortening the wavelength $\lambda$ in Eqn.(3), the flux pump's output current has been dramatically increased due to the effective increase of the magnetic gradient and coupling force [33]. Comparing with Giaever's "vortex-vortex" coupling in the DC transformer [2], this "magnet-vortices" coupling [11] is quite different, as it creates an induced current loop as the carrier for the coupled vortices. Despite the DC bias field could modulate the polarities of the coupled vortices in the current loop, and dramatically change the DC output voltage, however, the coupling force $F_{couple}$ is not affected by the DC field, based on Eqn.(7). That is to say, the DC output voltage can be freely modulated by the DC bias field, without change the coupled flux flow, the current distributions and the AC losses in the YBCO stator.

The above traits empowers travelling wave flux pump as a promising candidate as a powerful wireless DC power source for the next generation HTS magnet industry, with very large DC output [16], very accurate and versatile current control. By working the HTS magnet in a persistent current mode, the flux pumps also dramatically decrease the working power and cryogenic load comparing with conventional power sources with current leads. At the same time, the special electromagnetic behavior of travelling wave flux pump shed light to reconsider some fundamental problems in the understanding of superconducting electromagnetism.

## ACKNOWLEDGEMENTS

This work is supported by the National Natural Science Foundation of China under grant numbers 51877143, and the Science and Technology Project of Sichuan Province, China under grant number 2021YFS0088.

## Supplementary materials

**A. Numerical model.** The simulation was based on the *H*-formulation[1,2] with a finite-element method (FEM) software COMSOL Multiphysics. The geometry of the FEM model for the experiments is shown in Fig.6, which contains the linear-motor type flux pump, YBCO stator, superconducting load (a single YBCO wire), boundary. The Kirchhoff's current law is imposed between the YBCO stator and load to form a closed-loop, the same as in the simulation of HTS transformer-rectifier flux pump[2].

The results in Fig.2 are derived from this numerical model, while the internal flux flow for the three cases are derived from the central line marked as red in Fig.6.

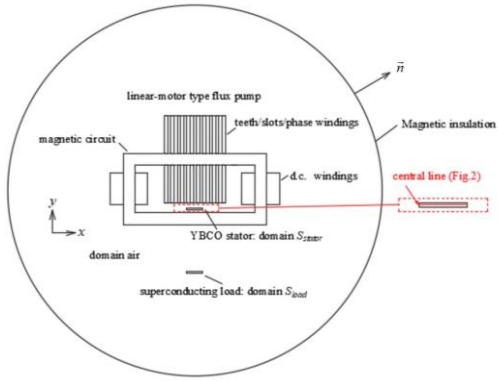

**Fig. 6 The FEM model based on *H*-formulation.** This model simulates the internal flux flow and current distributions inside the YBCO stator (Fig.2).

**B. Adaptive feedback control program on the output current.** For the accurate current control as shown in Fig.5, we first run the flux pump at its maximum output, that is, ensuring that $|B_{bias}| = B_a$. Then an adaptive feedback control program is used to control the DC windings of the flux pump to ensure that the pump current can be output at the target current $I_{pump}$ within an allowed fluctuation $\Delta I$. As shown in Fig.7, to accomplish the current control, the feedback program will track the difference between the output current and the preset current in real time, i.e., when the output current value is less than $I_{pump} - \Delta I/2$, the DC windings is turned on for charging, and when the current value is greater than $I_{pump} + \Delta I/2$, the DC windings is turned off to allow current decay.

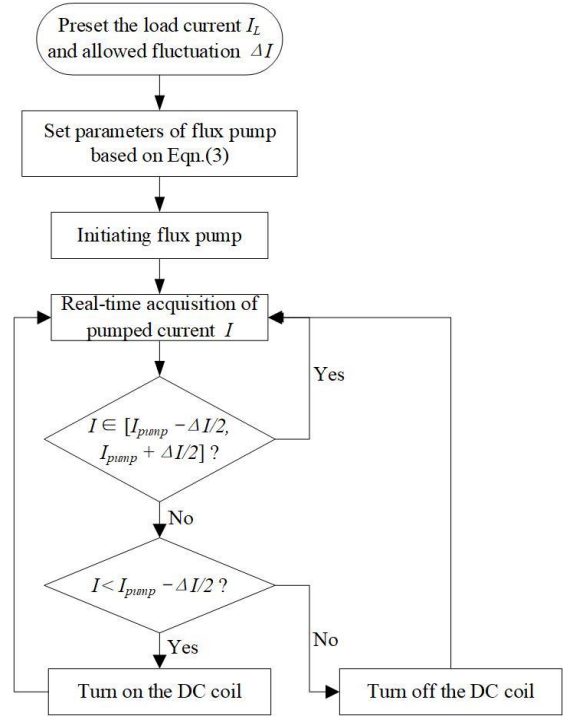

**Fig. 7 Adaptive control flow chart.** By comparing the difference between the output current and the target current, we perform a real-time feedback control of the output current, by controlling the on/off state of the DC coil.

**C. Experimental verifications on the directional relationship between $\vec{V}$, $\vec{B}$ and $\vec{v}$.** To verify that the DC output of travelling wave flux pump is a direct result of the motional *e.m.f.* expressed in Eqn.(8), six Hall sensors are installed on a printed circuit board (PCB), as shown in Fig.8(a). This Hall sensor array is inserted in the airgap of the linear-motor type flux pump, as shown in Fig.8(b), which measures the directions of $\vec{B}$ and $\vec{v}$ in the experiments Case (II) and (III). As shown in Fig.8(c), we have experimentally verified that the flux pump's DC output voltage $\vec{V}$ strictly follows the motional *e.m.f.* $\vec{E} = \vec{B} \times \vec{v}$ relationship based on right-hand rule.

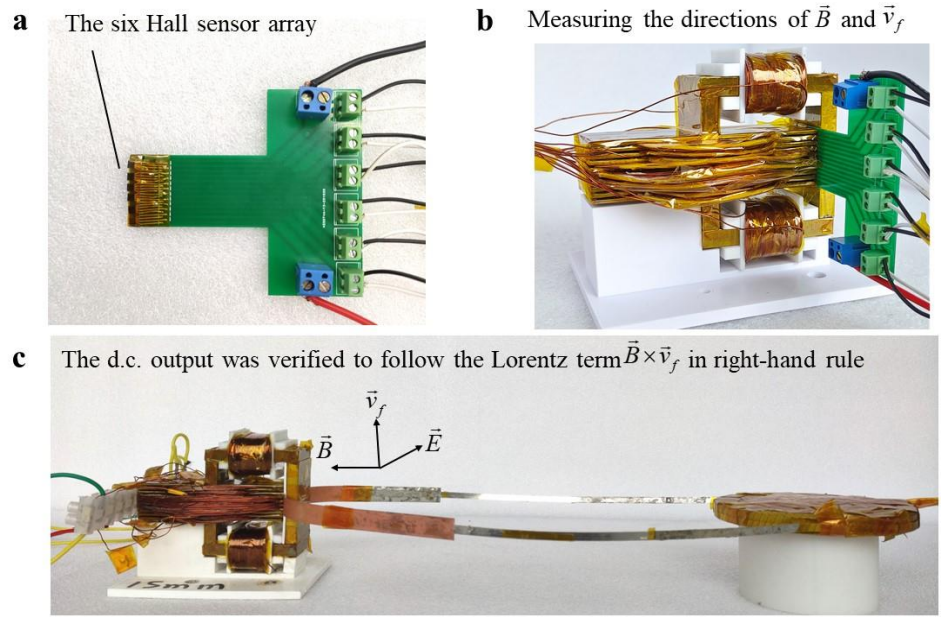

**Fig. 8 Experimental verifications on flux pump's DC output follows Eqn.(8). a** the six Hall sensors installed on the PCB. **b** the Hall sensor array was inserted in the airgap to measure the directions of $\vec{B}$ and $\vec{v}$ in the experiments Case (II) and (III). **c** the experiments verified that flux pump's DC output follows the motional *e.m.f.* expressed by Eqn.(8) based on right-hand rule.